\documentclass[sigconf]{acmart}

\acmConference{}{}{}
\acmBooktitle{}
\setcopyright{none}

\usepackage{stfloats}
\fnbelowfloat
\usepackage{algorithmic}
\usepackage{textcomp}
\usepackage{multirow}

\copyrightyear{2026}
\acmYear{2026}
\setcopyright{cc}
\setcctype{by}
\acmConference[NLBSE '26]{5th International Workshop on Natural Language-based Software Engineering }{April 12--18, 2026}{Rio de Janeiro, Brazil}
\acmBooktitle{5th International Workshop on Natural Language-based Software Engineering (NLBSE '26), April 12--18, 2026, Rio de Janeiro, Brazil}
\acmPrice{}
\acmDOI{10.1145/3786164.3794837}
\acmISBN{979-8-4007-2396-4/2026/04}

\begin{document}

\title{LoRA-MME: Multi-Model Ensemble of LoRA-Tuned Encoders for Code Comment Classification}

\author{Md Akib Haider}
\affiliation{%
  \institution{Islamic University of Technology}
  \country{Bangladesh}
}
\email{akibhaider@iut-dhaka.edu}

\author{Ahsan Bulbul}
\authornote{Both authors contributed equally.}
\affiliation{%
  \institution{Islamic University of Technology}
  \country{Bangladesh}
}
\email{ahsanbulbul@iut-dhaka.edu}

\author{Nafis Fuad Shahid}
\authornotemark[1]
\affiliation{%
  \institution{Islamic University of Technology}
  \country{Bangladesh}
}
\email{nafisfuad21@iut-dhaka.edu}

\author{Aimaan Ahmed}
\affiliation{%
  \institution{Islamic University of Technology}
  \country{Bangladesh}
}
\email{aimaanahmed@iut-dhaka.edu}

\author{Mohammad Ishrak Abedin}
\affiliation{%
  \institution{Islamic University of Technology}
  \country{Bangladesh}
}
\email{ishrakabedin@iut-dhaka.edu}
\begin{abstract}
Code comment classification is a critical task for automated software documentation and analysis. In the context of the NLBSE'26 Tool Competition, we present \textbf{LoRA-MME}, a Multi-Model Ensemble architecture utilizing Parameter-Efficient Fine-Tuning (PEFT). Our approach addresses the multi-label classification challenge across Java, Python, and Pharo by combining the strengths of four distinct transformer encoders: UniXcoder, CodeBERT, GraphCodeBERT, and CodeBERTa. By independently fine-tuning these models using Low-Rank Adaptation (LoRA) and aggregating their predictions via a learned weighted ensemble strategy, we maximize classification performance without the memory overhead of full model fine-tuning. Our model achieved an \textbf{F1 Weighted score of 0.7906} and a \textbf{Macro F1 of 0.6867} on the test set. However, the computational cost of the ensemble resulted in a final submission score of 41.20\%, highlighting the trade-off between semantic accuracy and inference efficiency. Our model and source code are publicly available at \href{https://huggingface.co/akibhaider/LoRA-MME-Code-Comment-Classifier}{Huggingface}.
\end{abstract}


\maketitle

\section{Introduction}

Code comments serve as a crucial bridge between source code and human understanding, facilitating software maintenance, comprehension, and evolution. As software systems grow in complexity, the ability to automatically categorize comments into semantic types—such as summary, usage, parameters, and deprecation warnings—becomes increasingly valuable for documentation generation, code search, and developer assistance tools.

Code comments carry distinct semantic meanings that can be systematically categorized~\cite{pascarella2017}. The code comment classification task was first formalized through introducing a taxonomy of comment types for Java software systems that form the basis of the NLBSE'26 Tool Competition~\cite{nlbse2026}.

Traditional approaches to this task often rely on handcrafted features or general-purpose sentence embeddings like Sentence-BERT~\cite{sbert}. While effective for general text, these models may not fully capture the unique characteristics of code-related natural language, which often contains technical terminology, API references, and code-like syntax. More recent approaches have leveraged code-specific pre-trained models, but the challenge of balancing classification accuracy with computational efficiency remains.

To address these challenges, we propose \textbf{LoRA-MME}, a hybrid architecture that combines the representational power of multiple code-specialized transformer encoders with the parameter efficiency of Low-Rank Adaptation (LoRA)~\cite{lora}. Our approach introduces three key strategies:

\begin{enumerate}
    \item \textbf{Independent LoRA Fine-Tuning:} We adapt four complementary code encoders—UniXcoder~\cite{guo-etal-2022-unixcoder}, CodeBERT~\cite{feng-etal-2020-codebert}, GraphCodeBERT~\cite{guo2021graphcodebertpretrainingcoderepresentations}, and CodeBERTa~\cite{husain2020codesearchnetchallengeevaluatingstate}—using LoRA, which injects trainable low-rank matrices into the attention layers while keeping the pre-trained weights frozen. This reduces trainable parameters to approximately 4.5\% per model (5.9M parameters), enabling efficient fine-tuning on consumer hardware.
    
    \item \textbf{Weighted Ensemble Learning:} Rather than simple probability averaging, we learn category-specific mixing weights that allow the ensemble to dynamically prioritize different encoders based on the comment type. For instance, GraphCodeBERT may be weighted relatively higher for data-flow related categories due to its pre-training on code structure.
    
    \item \textbf{Per-Category Threshold Optimization:} We optimize classification thresholds independently for each language-category pair on the validation set, addressing class imbalance and improving F1 scores across underrepresented categories.
\end{enumerate}

Despite strong test performance (Weighted F1: 0.7906, Macro F1: 0.6867), the ensemble's high computational cost limited the submission score to 41.20\%, motivating future work on knowledge distillation to improve efficiency.
\section{Related Work}

The systematic study of code comment classification was pioneered by Pascarella and Bacchelli~\cite{pascarella2017}, who proposed a taxonomy of comment types for Java systems and demonstrated that comments carry distinct semantic meanings amenable to automatic classification. Rani et al.~\cite{rani2021} extended this work to a multi-language setting encompassing Java, Python, and Pharo, establishing the taxonomies that form the basis of the current NLBSE challenge~\cite{nlbse2026}. Recent competition submissions have explored lightweight approaches; notably, Al-Kaswan et al.~\cite{AlKaswan2023} achieved competitive results using STACC, a SentenceTransformers-based method with SetFit~\cite{tunstall2022efficientfewshotlearningprompts} for few-shot learning, portraying that efficient sentence embedding approaches can perform well on this task. The competition baseline employs SetFit, which fine-tunes sentence transformers via contrastive learning~\cite{11029440}. While these methods offer computational efficiency, they do not fully leverage code-specific pre-trained knowledge. Our work addresses this gap by combining code-specialized encoders (CodeBERT~\cite{feng-etal-2020-codebert}, GraphCodeBERT~\cite{guo2021graphcodebertpretrainingcoderepresentations}, UniXcoder~\cite{guo-etal-2022-unixcoder}, CodeBERTa~\cite{husain2020codesearchnetchallengeevaluatingstate}) with LoRA~\cite{lora} for parameter-efficient fine-tuning, enabling an ensemble approach that would otherwise be memory-prohibitive.

\section{Architectural details}

\subsection{Dataset Overview} 
The code comment classification dataset comprises 9,361 sentences extracted from 1,733 comments across 20 open-source projects, augmented with source code context. The task is formulated as multi-label classification to address 487 multi-category instances. The data distribution follows language-specific taxonomies: \textbf{Java} (6,595 sentences) includes 7 categories: \textit{Summary, Pointer, Deprecation, Rational, Ownership, Usage, Expand}. \textbf{Python} (1,658 sentences) includes 5 categories: \textit{Summary, Parameters, Usage, DevelopmentNotes, Expand}. Finally, \textbf{Pharo} (1,108 sentences) comprises 6 categories: \textit{KeyImplementationPoints, Example, Responsibilities, Intent, KeyMessages, Collaborators}.

\subsection{Base Models}
We selected an ensemble of four models which are fine-tuned independently to ensure diverse feature extraction. Their outputs are then combined using a learned weighting mechanism:
\begin{itemize}
    \item \textbf{UniXcoder (Microsoft):} Utilized for its ability to handle cross-modal tasks and AST representations \cite{guo-etal-2022-unixcoder}.
    \item \textbf{CodeBERT (Microsoft):} Provides robust semantic alignment between natural language comments and code \cite{feng-etal-2020-codebert}.
    \item \textbf{GraphCodeBERT (Microsoft):} Incorporates semantic-level structure (data flow), which is crucial for categories like \textit{Pointer} and \textit{Usage} \cite{guo2021graphcodebertpretrainingcoderepresentations}.
    \item \textbf{CodeBERTa (Hugging Face):} A compact RoBERTa-based model pre-trained on code, offering complementary representations with lower computational overhead \cite{husain2020codesearchnetchallengeevaluatingstate}.
\end{itemize}

\subsection{LoRA Configuration}
We inject LoRA adapters into the `query', `key', `value', and `dense' layers of the attention mechanism. Based on our grid search, we utilized the following hyperparameters:
\begin{itemize}
    \item \textbf{Rank ($r$):} 16
    \item \textbf{Alpha ($\alpha$):} 32
    \item \textbf{Dropout:} 0.1
\end{itemize}
This configuration results in approximately 4.5\% trainable parameters per model (approx. 5.9M parameters), allowing for efficient fine-tuning on consumer hardware (RTX 3090).

\subsection{Ensemble Strategy}
\begin{figure}[t]
\centering
\includegraphics[width=\columnwidth]{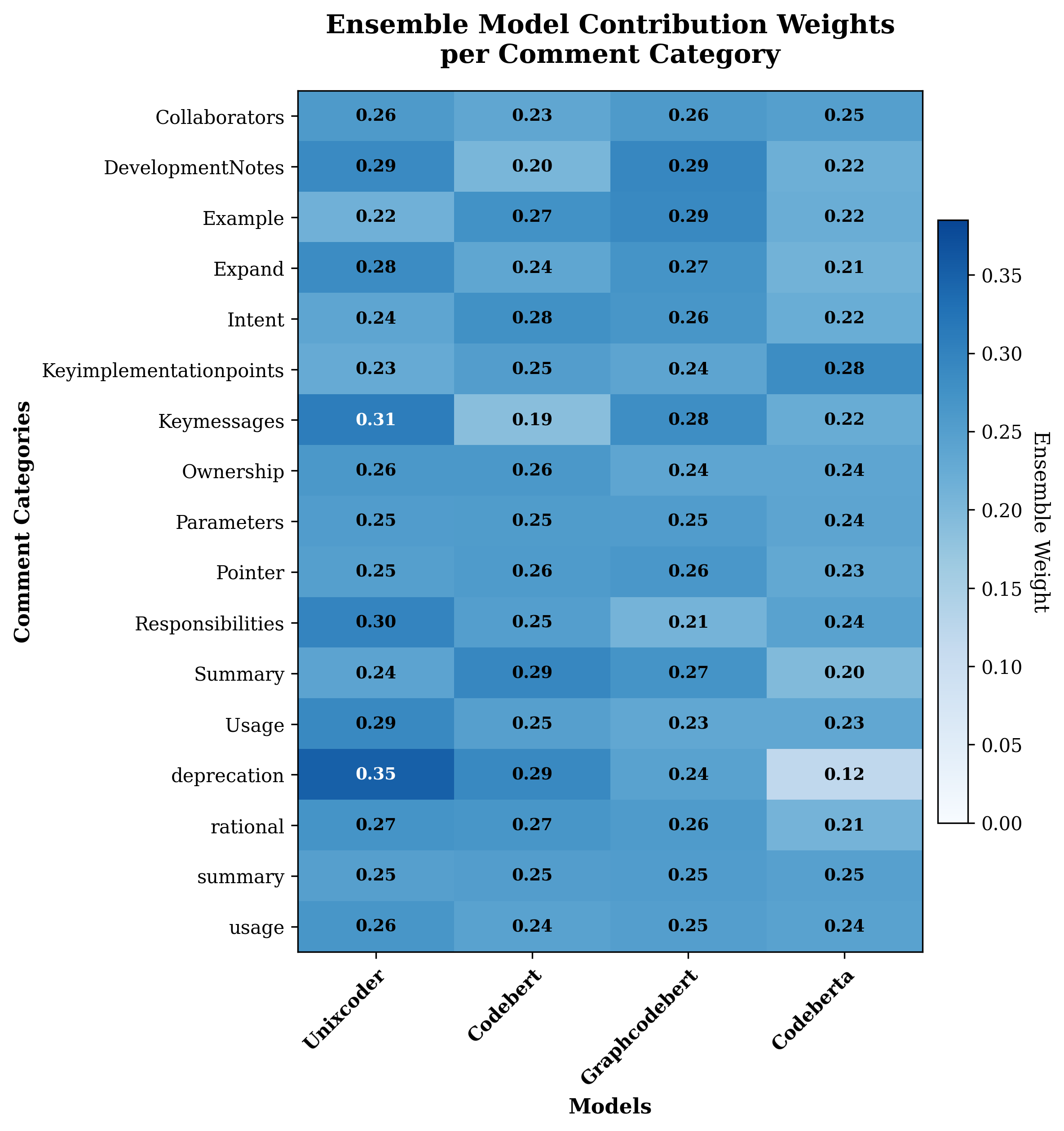}
\caption{Learned ensemble weights per category. Darker cells indicate higher contribution. The model dynamically assigns higher importance to specific encoders based on the category (e.g., UniXcoder for Pharo/Example).}
\Description{A heatmap showing the weight distribution of the 4 models across 17 categories.}
\label{fig:ensemble_weights}
\end{figure}

We apply a learned weighted ensemble strategy with softmax-like transformation to determine the final weights. For each category $c$, we learn a weight vector $W_c = [w_{1,c}, w_{2,c}, w_{3,c}, w_{4,c}]$ corresponding to the four models. The final probability is computed as:
\begin{equation}
P(c|x) = \sum_{m=1}^{4} w_{m,c} \cdot \sigma(z_{m,c})  
\end{equation}
Figure \ref{fig:ensemble_weights} illustrates the learned weights, showing how different models are prioritized for different categories.

\section{Methodology}

\begin{table*}[!b]
\centering
\caption{Per-Category Classification Performance (Test Set)}
\label{tab:results}
\begin{tabular}{llccc}
\toprule
\textbf{Language} & \textbf{Category} & \textbf{Precision} & \textbf{Recall} & \textbf{F1-Score} \\
\midrule
\multirow{7}{*}{\textbf{Java}} 
& Summary & 0.8184 & 0.9628 & 0.8848 \\
& Ownership & 0.8750 & 1.0000 & \textbf{0.9333} \\
& Expand & 0.4419 & 0.4810 & 0.4606 \\
& Usage & 0.9385 & 0.8271 & 0.8793 \\
& Pointer & 0.7792 & 0.9600 & 0.8602 \\
& Deprecation & 1.0000 & 0.7000 & 0.8235 \\
& Rational & 0.5000 & 0.2931 & 0.3696 \\
\midrule
\multirow{5}{*}{\textbf{Python}} 
& Usage & 0.8939 & 0.6484 & 0.7516 \\
& Parameters & 0.9032 & 0.6588 & 0.7619 \\
& DevelopmentNotes & 0.4583 & 0.3438 & 0.3929 \\
& Expand & 0.6452 & 0.3922 & 0.4878 \\
& Summary & 0.6753 & 0.8525 & 0.7536 \\
\midrule
\multirow{6}{*}{\textbf{Pharo}} 
& Key Impl. Points & 0.5600 & 0.5000 & 0.5283 \\
& Example & 0.9268 & 0.8539 & \textbf{0.8889} \\
& Responsibilities & 0.5714 & 0.7619 & 0.6531 \\
& Intent & 0.7692 & 0.9524 & 0.8511 \\
& Key Messages & 0.5882 & 0.6667 & 0.6250 \\
& Collaborators & 0.3333 & 0.7143 & 0.4545 \\
\midrule
\multicolumn{2}{l}{\textbf{Overall Macro Average}} & -- & -- & \textbf{0.6867} \\
\multicolumn{2}{l}{\textbf{Overall Weighted Average}} & -- & -- & \textbf{0.7906} \\
\bottomrule
\end{tabular}
\end{table*}

\subsection{Data Processing}
We identified and corrected text corruption where carets (\texttt{\textasciicircum}) appear instead of dot characters (\texttt{.}). As \texttt{\textasciicircum} is a valid return operator in Pharo, we correct caret corruption conservatively using context-based rules; for Java and Python, we replace all occurrences of \texttt{\textasciicircum}. Language-specific documentation patterns were retained: JavaDoc tags and common HTML elements (e.g., \texttt{<code>}) for Java; Sphinx-style tags (e.g., \texttt{:param}, \texttt{:return}) for Python; and Smalltalk-specific operators (e.g., \texttt{:=}, \texttt{\textasciicircum}, \texttt{\#}, \texttt{[ ]}, \texttt{|}) for Pharo. We implement a NEON-like heuristic to segment structured Pharo comments by splitting on line breaks and colon delimited headers, while preserving message-passing syntax by masking keyword selectors (e.g., \texttt{at:put:}). Case-aware text normalization (Camel, Pascal) was applied and version numbers were retained. High class imbalance (Java dominant) was handled during fine-tuning using focal loss with positive-class weighting~\cite{Lin_2017_ICCV}.

\subsection{Setup and Training}
LoRA-MME employs a multi-model ensemble combining UniXcoder, CodeBERT, GraphCodeBERT, and CodeBERTa. Each encoder is augmented with LoRA adapters~\cite{lora} ($r=16$, $\alpha=32$, dropout$=0.1$) targeting query, key, value, and dense layers, yielding approximately 5.9M trainable parameters (4.5\% of total).

Training uses Focal Loss ($\gamma=2.0$) to address class imbalance. Models are trained for 20 epochs with batch size 16, learning rate $2 \times 10^{-4}$, AdamW optimizer ($\epsilon=10^{-8}$, weight decay$=0.01$), and 10\% linear warmup. Best checkpoints are selected via validation macro-F1 with early stopping (patience$=3$).

\subsection{Threshold Optimization}

For multi-label classification, we optimize decision thresholds per (language, category) pair rather than using a fixed 0.5 threshold. A grid search over $t \in [0.1, 0.9]$ with step size 0.02 maximizes F1-score on validation data for each combination. This yields thresholds ranging from 0.28 to 0.85 (mean: 0.65), substantially improving classification performance over default thresholds.

\section{Results}

\subsection{Quantitative Analysis}
Classification performance was evaluated using Precision ($P_c$), Recall ($R_c$), and the harmonic mean $F_{1,c}$ for each category $c$, defined as:
\begin{equation}
P_c = \frac{ TP_c }{ TP_c + FP_c }, \quad
R_c = \frac{ TP_c }{ TP_c + FN_c }, \quad
F_{1,c} = 2 \cdot \frac{ P_c \cdot R_c }{ P_c + R_c }
\end{equation}

The final submission score was calculated using a composite formula that balances accuracy with inference latency ($T_{model}$) and computational cost ($G_{model}$):
\begin{equation}
\begin{aligned}
Score &= 0.6 \cdot \overline{F_1} 
+ 0.2 \cdot \max\left(\frac{T_{max} - T_{model}}{T_{max}}, 0\right) \\
&+ 0.2 \cdot \max\left(\frac{G_{max} - G_{model}}{G_{max}}, 0\right)
\end{aligned}
\end{equation}
where $\overline{F_1}$ represents the average F1 score across categories. To ensure comparability, runtime $T_{model}$ was measured on a standardized Google Colab T4 instance.

Our ensemble achieved an overall \textbf{F1 Macro score of 0.6867} and an \textbf{F1 Weighted score of 0.7906}. The detailed breakdown per category is presented in Table \ref{tab:results}.

Table \ref{tab:per_language} highlights that while Java performance remains robust ($F1=0.7445$), the model achieves significant gains in Python and Pharo compared to baselines.

\begin{table}[t]
\centering
\caption{Per-Language Aggregate Performance}
\label{tab:per_language}
\begin{tabular}{lccc}
\toprule
\textbf{Language} & \textbf{Macro F1} & \textbf{Baseline F1} & \textbf{$\Delta$} \\
\midrule
Java   & 0.7445 & 0.7306 & +0.0139 \\
Python & 0.6296 & 0.5820 & +0.0476 \\
Pharo  & 0.6668 & 0.6152 & +0.0516 \\
\midrule
\textbf{Overall} & \textbf{0.6867} & 0.6508 & \textbf{+0.0359} \\
\bottomrule
\end{tabular}
\end{table}

\begin{table}[t]
\centering
\caption{Threshold Optimization Ablation Study}
\label{tab:threshold_ablation}
\begin{tabular}{lcc}
\toprule
\textbf{Method} & \textbf{Macro F1} & \textbf{Weighted F1} \\
\midrule
Fixed Threshold (0.5)        & 0.6512 & 0.7654 \\
Per-Category Optimized       & \textbf{0.6867} & \textbf{0.7906} \\
\midrule
\textbf{Improvement}         & +0.0355 & +0.0252 \\
\bottomrule
\end{tabular}
\end{table}

\subsection{Performance vs. Efficiency}
The competition score is a composite of F1, Runtime, and GFLOPS.
\begin{itemize}
    \item \textbf{Average Runtime:} 45.13 ms/sample
    \item \textbf{Total GFLOPS:} $\approx$ 235,759.28
    \item \textbf{Submission Score:} 41.20\%
\end{itemize}

\begin{figure}[tbp]
\centering
\includegraphics[width=0.9\columnwidth]{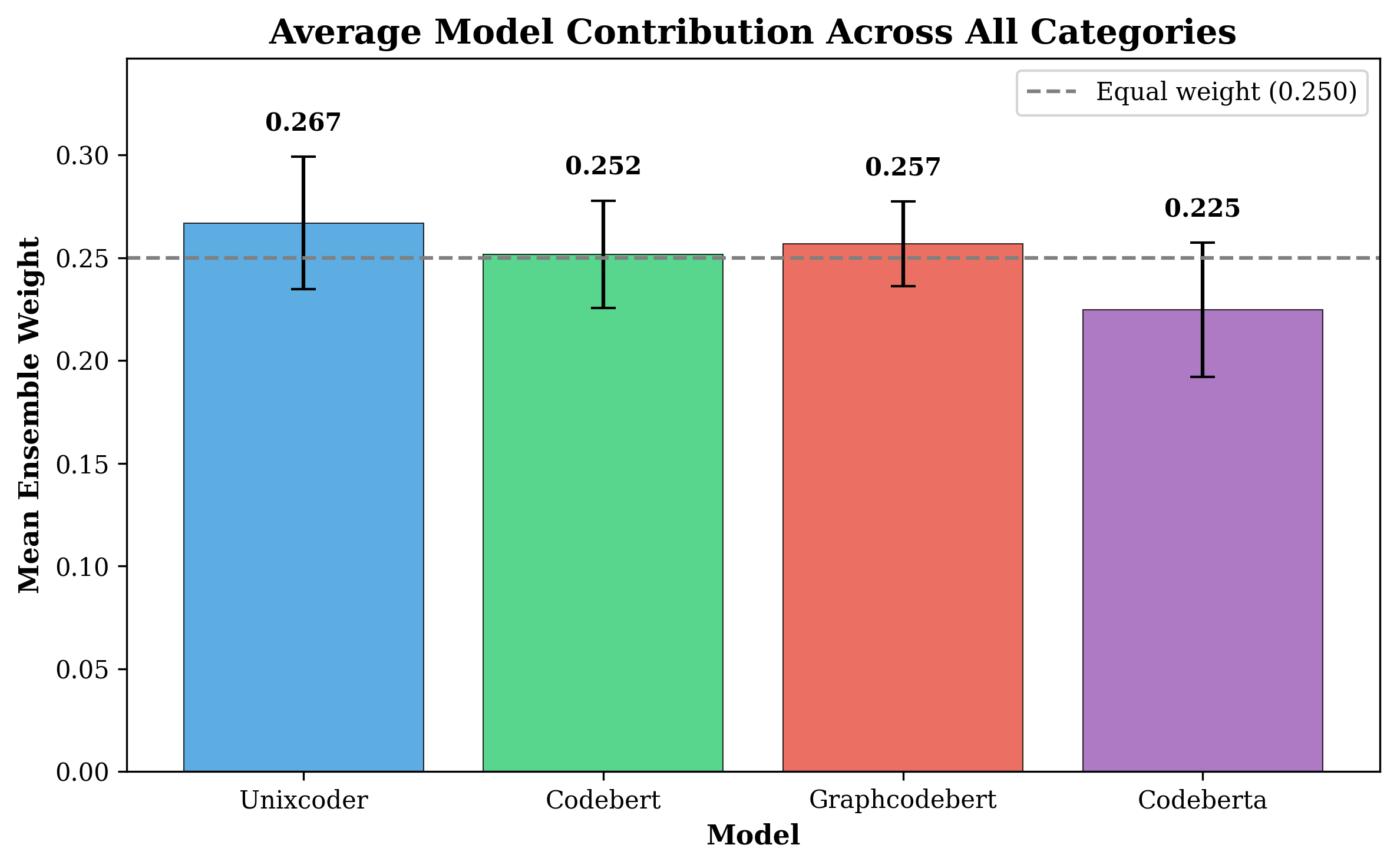}
\caption{Average Model Contribution Summary. The chart visualizes the aggregated importance of UniXcoder, CodeBERT, GraphCodeBERT, and CodeBERTa across all categories.}
\Description{A chart showing the relative contribution of each model to the ensemble.}
\label{fig:model_contrib}
\end{figure}

Figure \ref{fig:model_contrib} highlights the contribution of each model to the final ensemble. While the ensemble boosts semantic understanding, the computational cost is significant.

\begin{table}[t!]
\centering
\caption{Optimized Classification Thresholds Ablation Study (Illustrative Samples)}
\label{tab:thresholds}
\small
\begin{tabular}{llc}
\toprule
\textbf{Lang} & \textbf{Category} & \textbf{Threshold} \\
\midrule
Java & Summary & 0.49 \\
Java & Ownership & 0.28 \\
Java & Usage & 0.78 \\
Python & Usage & 0.78 \\
Python & Parameters & 0.85 \\
Pharo & Key Impl. Points & 0.74 \\
Pharo & Example & 0.70 \\
\bottomrule
\end{tabular}
\end{table}

\section{Conclusion}
We presented LoRA-MME, utilizing UniXcoder, CodeBERT, GraphCodeBERT, and CodeBERTa with LoRA adapters. The tool achieves strong classification results, particularly in detecting \textit{Ownership} and \textit{Usage} across languages. Our per-category threshold optimization strategy contributed an improvement of +0.0355 in Macro F1 over the fixed threshold baseline. Future work will focus on knowledge distillation—training a single student model to mimic this ensemble—to improve the submission score by reducing GFLOPS.


\bibliographystyle{ACM-Reference-Format}
\bibliography{references}

\end{document}